\documentclass[9pt,twocolumn,twoside]{opticajnl}

\journal{optica} 

\setboolean{shortarticle}{true}

\usepackage{comment}
\usepackage{lineno}

\newcommand{\p}{\partial}
\newcommand{\wh}{\widehat}

\newcommand{\ep}{\varepsilon}
\newcommand{\vta}{\vartheta}
\newcommand{\om}{\omega}

\newcommand{\al}{\alpha}

\newcommand{\cE}{{\cal E}}

\newcommand{\cN}{{\cal N}}
\newcommand{\cP}{{\cal P}}

\newcommand{\cF}{{\cal F}}

\newcommand{\cW}{{\cal W}}

\newcommand{\be}{\begin{equation}}                                       
	\newcommand{\ee}{\end{equation}}
\newcommand{\ba}{\begin{eqnarray}}
	\newcommand{\ea}{\end{eqnarray}}
\newcommand{\bref}[1]{(\ref{#1})}

\newcommand{\bi}[1]{\bibitem{#1}}\newcommand{\lab}[1]{\label{#1}}

\newcommand{\bsub}{\begin{linenomath}\begin{subequations}}                      
		\newcommand{\esub}{\end{subequations}\end{linenomath}}

\title{Carrier-resolved real-field theory of multi-octave frequency combs}

\author[1,2]{Danila N. Puzyrev}
\author[1,2,*]{Dmitry V. Skryabin}

\affil[1]{Department of Physics, University of Bath, BA2 7AY, UK}
\affil[2]{Centre for Photonics and Photonic Materials, University of Bath, BA2 7AY, UK}

\affil[*]{Corresponding author: d.v.skryabin@bath.ac.uk}

\begin{abstract}
	Optical frequency combs are pillars of precision spectroscopy, and their microresonator realisation serves applications where miniaturisation and large tooth separation are important.
Microresonator combs cover an enormous range of time scales varying from the femtosecond periods of optical oscillations to milliseconds corresponding to the kilohertz linewidth of the comb teeth. Here, we develop and implement the carrier-resolved real-field model for multi-octave frequency combs, which allows for near ab-initio capture of all the time scales involved. As an example, we consider a microresonator that has a mix of second and third-order nonlinearities and uses periodic poling. By applying the real-field approach, we demonstrate how to surpass traditional limitations and model the spectral broadening and soliton mode-locking across three optical octaves.
\end{abstract}

\setboolean{displaycopyright}{true}

\begin{document}
	
	\maketitle
Nonlinear optical frequency conversion was observed at the beginning of the laser era~\cite{bass}, and since then, it has been one of the most active areas of applied and fundamental research~\cite{boyd}. The development of high power femtosecond and attosecond laser systems has opened numerous opportunities to use nonlinear processes for the generation of electromagnetic spectra between XUV and THz frequency bands~\cite{road, road2, zhel}. This has created a demand for an efficient theoretical and computational framework to model ultra-broad spectra, which could not be described by the mid-20th century theoretical tools of nonlinear optics. A variety of generalized envelope equations~\cite{kosareva, brabec, skr, dudl, chi2a}, and a class of carrier-resolved models, also known as unidirectional pulse-propagation equations (UPPE)~\cite{kolesik1, guide, gusakov, trav, bang}, have met this demand.

Optical frequency combs and associated soliton pulses in microresonators have recently become a fast-growing area and a contender in the generation of temporally ultra-short and spectrally ultra-broadband waveforms~\cite{rev0,rev1}. The advantages of microresonators include a small footprint, small operational powers, and dispersion engineering used for spectral tuning of the generated combs. Microresonator comb spectra spanning across multiple octaves from mid-infrared to visible have been observed in~\cite{old, jap, ing, china, kart}. Contrary to filamentation and supercontinuum experiments using high-energy ultra-short laser pulses, a typical microresonator experiment requires milliwatts of CW power, while short pulses are generated inside the resonator.

A numerical model well suited to describe comb generation in microresonators can be formulated using a coupled-mode theory~\cite{herr,chembo,osac}. This approach leads to a set of ordinary differential equations for amplitudes of the resonator modes evolving in time. Fourier transforming known coupled-mode equations to physical space, one finds the driven-damped nonlinear Schr\"odinger equation, which is known in the resonator context as the Lugiato-Lefever equation, or other envelope models~\cite{osac,lug1,lug2,chi2b,josab}. Approximations underpinning the envelope models rely on the selection of a reference carrier frequency and wave-number and their subtraction from the rest of the spectrum. 
	
Our aim here is to bypass these approximations and present a carrier-resolved real-field theory for multi-octave frequency combs. We want this theory to capture desirable features of the carrier-resolved models known for propagation in bulk crystals, gases, and  waveguides~\cite{guide,gusakov,trav,bang}. One such property is using the real-valued electric field, $\cE$, oscillating with the required spectrum of electromagnetic frequencies. The real field can be conveniently used to express the second, $\cE^2$, and third, $\cE^3$, order effects without omitting 
any of nonlinear terms, which is an unavoidable shortcoming 
of the envelope based methods~\cite{josab,chi2b}.  More generally, the carrier-resolved approach has not been yet formulated for a broad class of the resonator systems where pump, gain and dissipation are intrinsic for device operation~\cite{grelu,tangn,pup2,marandi}, and where current trend towards the generation of increasingly extreme waveforms~\cite{road,road2} requires development of novel theoretical methods beyond the envelope Haus-master and Ginzburg-Landau equations~\cite{grelu}.

To develop the carrier-resolved theory of ring microresonators, we proceed with the initial steps of the formalism described in~\cite{josab}. This formalism seeks the electric field components, $\cE_\alpha$, obeying Maxwell equations as a superposition of the resonator modes $F_{\al j}(r,z)e^{ij\vta- i\om_jt}$, where $j$ is an azimuthal mode number, $\alpha$ can be either $x$, $y$ or $z$,
$\vta=[0,2\pi)$ is the polar angle and $r$ is the distance in the $(x,y)$-plane, $z$ axis is perpendicular to the resonator plane,
and $\om_j$ are the resonance frequencies.
We account only per one mode for every $j$, and assume that this mode is dominated by one polarization direction which allows us to omit the subscript $\al$, 
\be
\cE =
\sum\limits_{j=j_{\min}}^{j_{\max}}b_{j}B_{j}(t) F_{j}(r,z)e^{ij\vta-i\om_{j} t}+c.c.,
\lab{field}
\ee
and use  scalar approximation for nonlinear polarization.
Here, $B_j(t)$ are the unknown modal amplitudes. If $\cE$ and $b_jB_j$ are measured in  V/m, then choosing $b_j^2=2/\epsilon_0cn_jS_j$, makes  $|B_j|^2$ to have units of Watts, where $n_j$ is the refractive index and $S_j$ is the transverse mode area.  
Ultimately, our approach is superior over the envelope and prior coupled-mode methods because it explicitly uses a continuous span of the resonator modes extending over two, three or more octaves, and true real fields 
to evaluate the nonlinear terms, which thereby captures all possible nonlinear 
wave-mixing processes. 
	
In a resonator, every mode experiences damping with the rate $\kappa_j$, while the comb generation is achieved by pumping one of the modes, e.g., $j=M$, with laser light having frequency $\om_p$ close to $\om_M$. Substituting Eq.~\bref{field} to Maxwell equations, we find a set of equations for modal amplitudes, 
\be 
\begin{split}
&	i \p_tB_{j}=-i\tfrac{1}{2}\kappa_j\big( B_j-\wh\delta_{j,M}\cP
e^{i\delta t}\big)\\
&-\frac{\om_j}{2 n_j^2b_j} 
\frac{e^{i\om_j t}\iiint F_j e^{-ij\vta} \cN(\cE)
r{\rm d}r{\rm d}z{\rm d}\vta}
{\iiint F_j^2 r{\rm d}r{\rm d}z{\rm d}\vta }.     
\end{split}
\lab{mod2b}
\ee
Here, $\delta=\om_M-\om_p$ is the frequency detuning, and $\wh\delta_{j,M}=1$ for $j=M$ and is zero otherwise. $\cP=\sqrt{\cF\eta \cW/\pi}$ is the pump parameter, where $\cF$ is finesse, $\cW$ is laser power at the coupling point and $\eta$ is coupling coefficient. 

\begin{figure}[t]
	\centering{\includegraphics[width=0.45\textwidth]{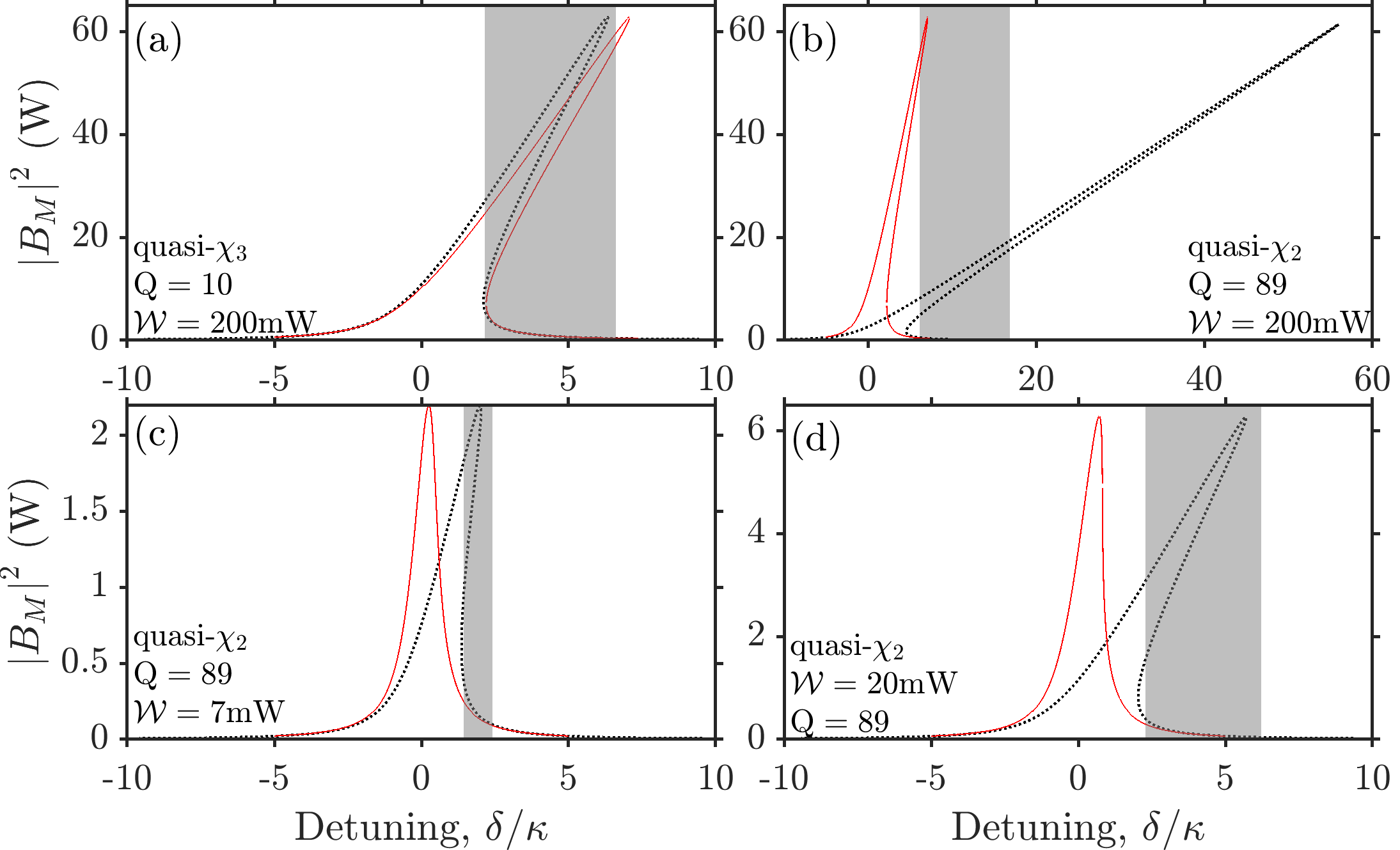}}
	\caption{Black (red) lines show power of the pump mode, $|B_M|^2$, vs detuning, $\delta$, when  $\tilde\chi^{(2)}+\chi^{(3)}$ (only $\chi^{(3)}$) nonlinearities are  accounted for.		
		$Q$ is the number of poling periods, and $\cW$ is laser power.  Shading shows intervals investigated for frequency comb generation using the carrier resolved model, Eq.~\bref{super}. Nonlinear parameters are $\gamma_{2,2M+Q}/2\pi=8.39$GHz/$\sqrt{\rm W}$ and $\gamma_{3,M}/2\pi=3.79$MHz/W, linewidth is $\kappa/2\pi=100$MHz, coupling parameter and finesse are $\eta=0.5$ and $\cF=3940$.}
	\label{fig1}
\end{figure}

\begin{figure}[t]
	\centering{\includegraphics[width=0.48\textwidth]{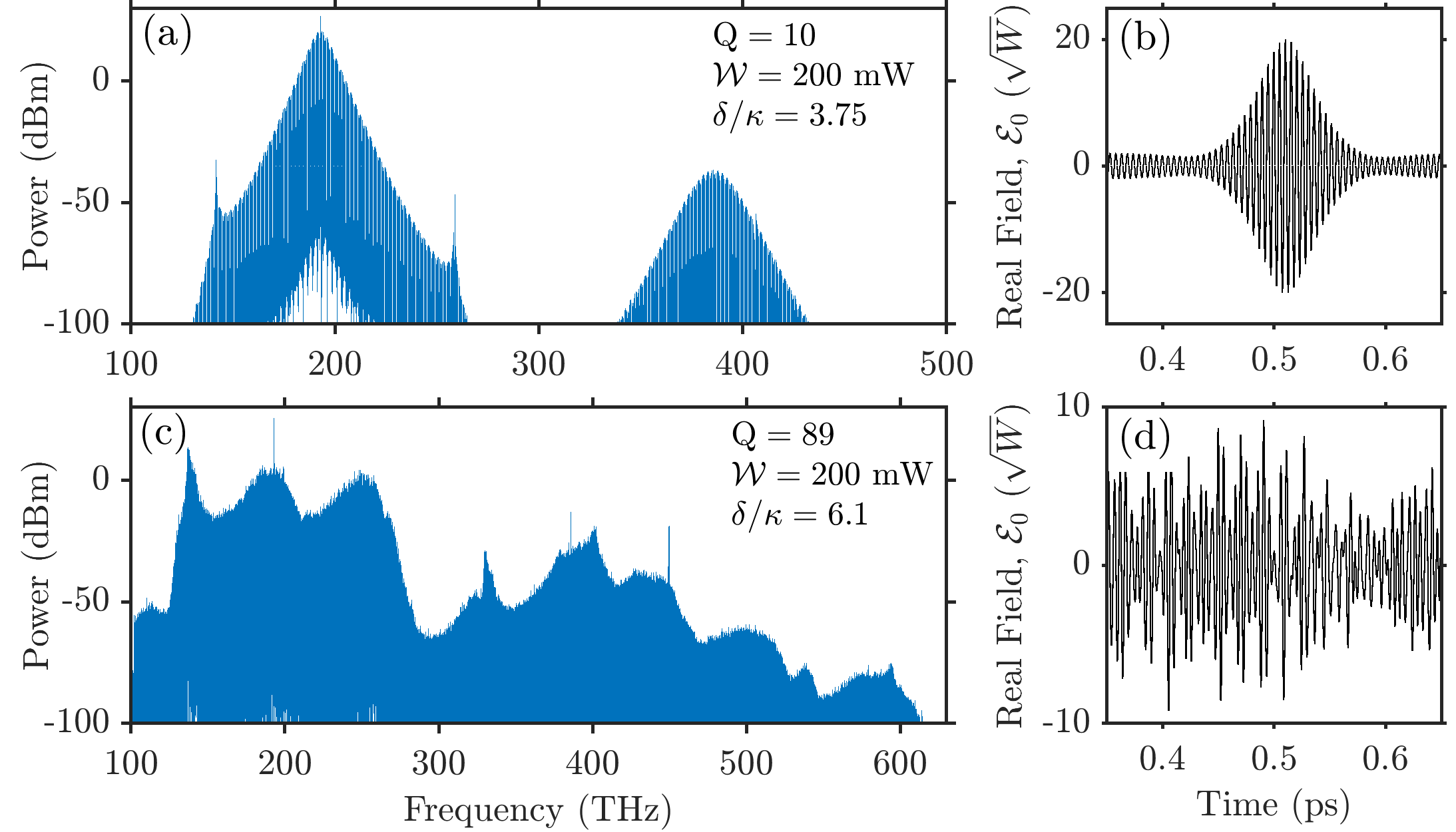}}
	\caption{ (a,b) Spectrum (Fourier transform, $t\to\om$, of $\cE_0(t,\vta=0)$) and real field profile, $\cE_0(t,\vta=0)$, of the Kerr soliton generated in strongly mismatched resonator, $Q=10$, $M=404$, $\ep_2/2\pi=30$THz.  (c,d) Three-octave spectrum corresponding to the randomly fluctuating electric field generated in quasi-phase-matched resonator, $Q=89$, $M=404$, $\ep_2/2\pi=-445$GHz, $j_{\min}=147$, $j_{\max}=1478$. Pump is  $\cW=200$mW in both cases. Total integration time is $191$ns in (a) and $143$ns in (b). Spectrum was computed over last $40$ns in (a) and $95$ns in (b). Round trip time is $2.538$ps. }
	\label{fig2}
\end{figure}

$\cN$ is the nonlinear part of material response, which in our case includes second-, $\tilde\chi^{(2)}$, and third-order, $\chi^{(3)}$, nonlinear susceptibilities, 
\be 
\cN(\cE)=\tilde\chi^{(2)}\cE^2+\chi^{(3)}\cE^3.
\lab{pr}\ee 
Second-order susceptibility can vary in $\vta$ if periodic poling is applied  for quasi-phase-matching, i.e., $\tilde\chi^{(2)}(\vta)=\tilde\chi^{(2)}(\vta+2\pi/Q)$, where $Q$ is the number of periods along the ring~\cite{tang1},
\be
\tilde\chi^{(2)}(\vta)=\chi^{(2)}g(\vta)=\chi^{(2)}
\sum_{m}g_me^{-imQ\vta},~m=\pm 1,\pm 2,\dots,
\lab{c2}
\ee
and $g(\vta)$ varies between $+1$ and $-1$. The no-poling case corresponds to 
$g(\vta)=1$. Quasi-phase-matching for the modes $M$ and $2M+Q$ is achieved, when $Q$
is selected to minimize the frequency mismatch parameter,
\be
\ep_2=2\om_M-\om_{2M+Q}.\lab{ep2}\ee

Detailed derivation of Eq.~\bref{mod2b} can be found in~\cite{josab}, see Eq.~(3.2) and Eq.~(6.1). Equation~\bref{mod2b} is also structurally the same as Eq.~(24) in~\cite{chembo}, where $\cN$ was the Kerr part of refractive index.	
A standard way to simplify and then solve Eq.~\bref{mod2b} is to introduce several groups of modes centred around the pump and its harmonics and simplify nonlinear terms  by limiting nonlinear wave-mixing processes to the ones that resonantly excite selected modes, which was also the approach used in~\cite{josab}. However, this method is compromised when the individual modal groups approach the octave width and omitted non-resonant nonlinear terms start playing a role. 

\begin{figure*}[t]
	\centering{\includegraphics[width=0.9\textwidth]{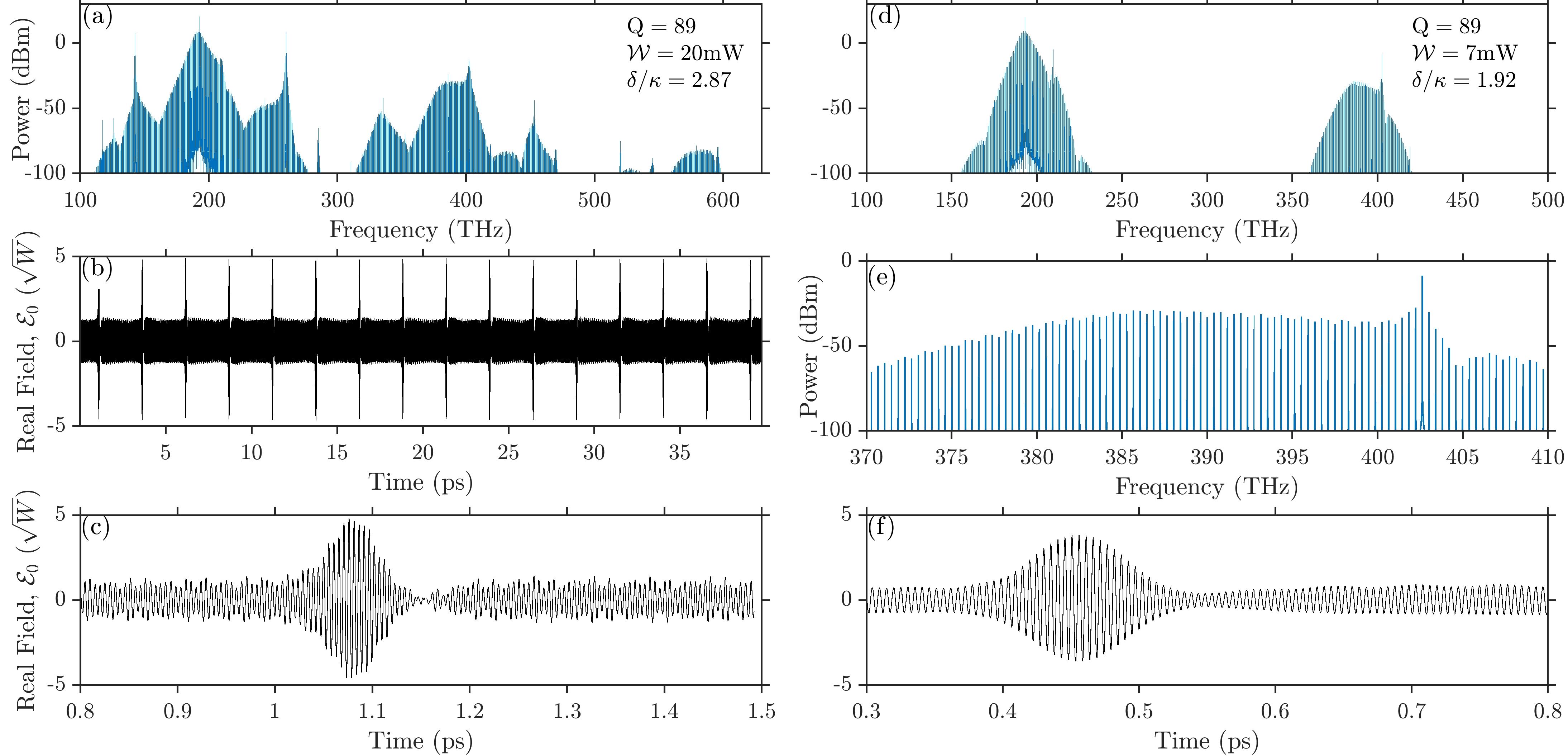}}
	\caption{$\chi^{(2)}$ solitons generated in quasi-phase-matched resonator, $Q=89$,  $M=405$, $\ep_2/2\pi=-445$GHz, for relatively small powers. (a-c) Spectrum, section of the soliton train and real field profile of a single soliton for $\cW=20$mW.   (d-f) Spectrum, its zoomed interval showing the resolved comb lines and real field profile of a single soliton for $\cW=7$mW. Total integration time is $190$ns in (a) and $143$ns in (b). Spectrum was computed over last $80$ns in (a) and $88$ns in (d). Round trip time is $2.538$ps.}
	\label{fig3}
\end{figure*}

Equation \bref{mod2b} is free from the above limitation. It treats all the modes on equal footing and accounts for all nonlinear wave-mixing processes by dealing with $\cE^2$ and $\cE^3$ without simplifications.
However, substituting Eq.~\bref{field} to the right-hand side of Eq.~\bref{mod2b}  yields tens of thousands of nonlinear coefficients varying with $j$ within around one order of magnitude, and the equations appear to be entirely not practical even before numerical integration in time is approached. To derive a model with the practical number of nonlinear coefficients and, at the same time, to retain all terms inside $\cE^2$ and $\cE^3$, we make a simplification that for every $j$ there is the same coefficient for all nonlinear mixing terms entering the respective equation, but the coefficients vary with $j$.
This is achieved by assuming $\iiint \cE^3F_je^{-ij\vta}r{\rm d}r{\rm d}z{\rm d}\vta\approx$ 
	$b_j^3\iint F_j^4r{\rm d}r{\rm d}z\int_0^{2\pi}\cE_0^3e^{-ij\vta}{\rm d}\vta$,
	and $\iiint g(\vta)\cE^2F_je^{-ij\vta}r{\rm d}r{\rm d}z{\rm d}\vta\approx$
	$b_j^2\iint F_j^3r{\rm d}r{\rm d}z\int_0^{2\pi}g(\vta)\cE_0^2e^{-ij\vta}{\rm d}\vta$,
	where
	\be \cE_0=
	\sum_{j=j_{\min}}^{j_{\max}}B_{j}e^{ij\vta-i\om_{j}t}+c.c.,
	\lab{super0}\ee
	is the real-valued carrier-resolved field, where the transverse profile of the modes has been factored out. Equations~\bref{mod2b} are then reduced to 
	\be 
	\begin{split}
		i \p_tB_{j}
		=&-\tfrac{1}{2}\kappa_j\big( B_j-\wh\delta_{j,M}\cP
		e^{i\delta t}\big)\\
		& -\gamma_{2,j}\int_0^{2\pi}
		g(\vta)\cE_0^2 e^{i\om_jt-ij\vta}\frac{{\rm d}\vta}{2\pi}\\
		& -\gamma_{3,j}\int_0^{2\pi} \cE_0^3 e^{i\om_jt-ij\vta}\frac{{\rm d}\vta}{2\pi}, 
	\end{split}
	\lab{super}
	\ee
which retain all possible nonlinear wave-mixing processes and oscillations at the carrier frequencies across all the relevant spectrum.
	Here, 
	\be
	\begin{split}
		&	\gamma_{3,j}=\frac{\om_jb_j^2\pi}{n_j^2}
		\frac{\iint \chi^{(3)} F_j^4r{\rm d}r{\rm d}z}{\iint F_j^2r{\rm d}r{\rm d}z},~[\gamma_{3,j}/2\pi]=\frac{\rm Hz}{\rm W},\\
		& \gamma_{2,j}=\frac{\om_jb_j\pi}{n_j^2}
		\frac{\iint \chi^{(2)} F_j^3r{\rm d}r{\rm d}z}{\iint F_j^2r{\rm d}r{\rm d}z},~[\gamma_{2,j}/2\pi]=\frac{\rm Hz}{\rm W^{1/2}},
	\end{split}
	\lab{gamma}
	\ee
	are nonlinear coefficients.
Numerical data presented in the main text 
further assume that $\gamma_{3,j}=\gamma_{3,M}$ and  $\gamma_{2,j}=\gamma_{2,2M+Q}$, 
while the spectra accounting for dispersion of $\gamma_{2,j}$, $\gamma_{3,j}$ are included in~\cite{data} for comparison.

Equation~\bref{super} is our main result. Structurally, it can be matched with the carrier-resolved pulse propagation models, see, e.g.,  Eq.~(85) in \cite{guide} and references therein. Apart from the pump and loss terms, a difference to be noted is that Eq.~\bref{super} describes the evolution of spatial harmonics (resonator modes) in time. At the same time, aforementioned carrier-resolved~\cite{guide} and multi-octave envelope models~\cite{dudl,chi2a,chi2b} deal with the evolution of discretized frequency spectrum in the propagation coordinate. Hence, unlike the above references, to reconstruct the frequency spectrum, we need to store the data in time and then Fourier transform them to frequency, $t\rightarrow \om$, as post-processing. An advantage of our method for describing resonator systems generating narrow spectral features is that the frequency grid is not preset, and frequencies are captured as generated.

As a practical example, we consider a thin-film lithium niobate (LN) microresonator with a geometry close to implemented in~\cite{tang2}. The ridge width at the top surface is 1.45$\mu$m, ridge height is $360$nm, wall angle is $60^o$, total LN thickness is $560$nm and the internal radius is 50$\mu$m, so the repetition rate at $\om_M/2\pi\approx 193$THz is $394$GHz.
Experimental measurements demonstrated that such geometry without periodic poling and pumped with the $\sim$100mW of power produces 100THz wide Kerr frequency combs with two strong dispersive waves emitted close to 150 and 250THz~\cite{tang2}. For the data sets with $\om_j$, $\gamma_{2,j}$, $\gamma_{3,j}$, and $g(\vta)$, see~\cite{data}.

The nonlinear part of Eq.~\bref{super} is formulated in a pseudo-spectral form~\cite{guide}, i.e., it requires going back to physical space to compute $\cE_0$. Fully spectral representation is also possible but computationally less efficient~\cite{josab}. In solving Eq.~\bref{super} numerically, the integrals and transformation from the real field, $\cE_0$, to the mode amplitudes, $B_{j}$, and back, $\vta\leftrightarrow j$, were implemented using FFT. The evolution in time was computed using the third-order Runge-Kutta method. The number of modes was set to embrace the third harmonic.
In that case, the maximal frequency of oscillations produced by the $\gamma_{3,j}$-terms is well estimated by $\ep_3=3\om_M-\om_{3M}$. $\ep_3$ is zero in the absence of dispersion (equidistant resonator modes), revealing that the maximal suitable time step is determined not by the optical frequency but by the resonator dispersion, which is enhanced in integrated photonics due to tight structural confinement. In our case, $|\ep_3|/2\pi\approx 75$THz. In practice, stability and convergence of numerical code were achieved for steps $\le\pi/4|\ep_3|\approx 1.7$fs. If the third-harmonic modes are set to zero, then the near-phase-matched resonators can be integrated using time steps $\sim 100$fs, determined by the minimal frequencies produced by the 
$\gamma_{2,j}$-terms, $|\ep_2|/2\pi$, see Eq.~\bref{ep2}. Setting second and third harmonics to zero increases permitted steps to above $1$ps.

In our resonator, modes $M=404,405$ ($\om_M/2\pi\approx 193$THz) are  quasi-phase-matched with $2M+Q$ for $Q=87,88,89$, 
so that $|\ep_2|$ reaches its minimum, where it becomes order of $100$GHz. 
We start presenting our simulation results from the strongly mismatched case $Q=10$. Bistable response of the pumped resonance for power $\cW=200$mW, which includes (black line) and excludes (red line) $\chi^{(2)}$ nonlinearity is shown in Fig.~\ref{fig1}(a). 
One can see that this arrangement corresponds to the quasi-Kerr regime since second-order nonlinearity makes little impact. Therefore, we expect to find Kerr-soliton combs with two dispersive waves as were observed in~\cite{tang2}. The shaded area in Fig.~\ref{fig1}(a) shows an interval of the pump detuning where we have found solitons and breathers by solving Eq.~\bref{super}. A typical spectrum and carrier-resolved soliton pulse are shown in Figs.~\ref{fig2}(a), (b).

Selecting $Q=89$, we bring the resonator to quasi-phase-matching. Figures~\ref{fig1}(b)-\ref{fig1}(d) show the respective bistable curves for $\cW=200$mW, $20$mW and $7$mW of pump power. For all three of them, tilts of the resonance happen dominantly through the $\chi^{(2)}$ nonlinearity and the soliton and comb generation in the shaded intervals of detunings were found outside the Kerr-only resonance (red line). Spectra computed in the $200$mW case, continuously extend across three octaves, see Fig.~\ref{fig2}(c). They correspond to the incoherent randomly varying fields, which carrier-resolved form is shown in Fig.~\ref{fig2}(d).

Reducing power to $20$mW and then to $7$mW we 
have observed formation of $\chi^{(2)}$ solitons and breathers, see Fig.~\ref{fig3}. For higher powers, these solitons are surrounded by strong dispersive waves emitted around the pump and second-harmonic. Dispersive wave emission is greatly reduced for powers less than $10$mW, where the soliton combs become spectrally narrow. 
Spectral resolution in our approach is set by the inverse integration time, which was $88$ns for Fig.~\ref{fig3} providing $11$MHz resolution.
A section of the recorded pulse train, zoomed-on spectra, and real-field profiles of individual solitons are also shown.

In summary: We developed the real-field carrier-resolved model to describe optical frequency comb generation in ring microresonators.
Our model is capable of reproducing frequency combs with the multi-octave spectral width, while simultaneously accounting for every mode physically present in the resonator and all nonlinear wave-mixing terms entering the nonlinear polarization specified via the real-valued electric field, see Eq.~\bref{pr}.  
For example, if we consider $\chi^{(2)}$ nonlinearity and two envelope functions $A_1$ and $A_2$ centred around pump and its second-harmonic,  the opening of the second-order nonlinear polarization  
$\chi^{(2)}(A_1e^{i\phi_1}+A_2 e^{i\phi_2}+c.c.)^2$ yields a combination of ten terms.
Here, $\phi_{1,2}$ are the  phases oscillating with optical frequencies.
Omission of any term fails to be justifiable if the spectra of $A_{1,2}$ approach an octave width, while our method is free from this constraint and solves the problem for nonlinearities of arbitrary orders.  Overall, our results open an avenue for theoretical research into a growing family of resonator systems requiring carrier-resolved multi-octave modelling of ultrashort pulses~\cite{pup2,marandi,stab}.

	\begin{backmatter}
		\bmsection{Acknowledgments} 	Authors acknowledge financial support from the  Royal Society (SIF/R2/222029) and thank A. Villois for discussions. 
		\bmsection{Disclosures} The authors declare no conflicts of interest.
		\bmsection{Data Availability Statement}   Data included in this study are openly available~\cite{data}.
	\bmsection{Author contributions} 
	DNP - coding and data acquisition. 	DVS - concept and theory. 
	\end{backmatter}

\end{document}